\documentclass[sigconf,screen]{acmart}
\usepackage{balance}
\usepackage{flushend}
\usepackage{graphicx}

\copyrightyear{2026}
\acmYear{2026}
\setcopyright{cc}
\setcctype{by}
\acmConference[ICSE-Companion '26]{2026 IEEE/ACM 48th International Conference on Software Engineering}{April 12--18, 2026}{Rio de Janeiro, Brazil}
\acmBooktitle{2026 IEEE/ACM 48th International Conference on Software Engineering (ICSE-Companion '26), April 12--18, 2026, Rio de Janeiro, Brazil}
\acmPrice{}
\acmDOI{10.1145/3774748.3795874}
\acmISBN{979-8-4007-2296-7/2026/04}

\begin{document}
\title{Sovereign-by-Design\\A Reference Architecture for AI and Blockchain Enabled Systems}

\author{Matteo Esposito}
\orcid{0000-0002-8451-3668}
\affiliation{
\institution{University of Oulu}
  \city{Oulu}
  \country{Finland}
}
\email{matteo.esposito@oulu.fi}

\author{Lodovica Marchesi}
\affiliation{
\institution{University of Cagliari}
  \city{Cagliari}
\country{Italy}}
\orcid{0000-0002-0627-5043}
\email{lodovica.marchesi@unica.it}

\author{Roberto Tonelli}
\affiliation{
\institution{University of Cagliari}
  \city{Cagliari}
\country{Italy}}
\orcid{0000-0002-9090-7698}
\email{roberto.tonelli@unica.it}

\author{Valentina Lenarduzzi}
\orcid{0000-0003-0511-5133}
\affiliation{
  \institution{University of Southern Denmark, Vejle}
  \city{Vejle}
  \country{Denmark}
}
\email{lenarduzzi@imada.sdu.dk}


\begin{abstract}
Digital sovereignty has emerged as a central concern for modern software-intensive systems, driven by the dominance of non-sovereign cloud infrastructures, the rapid adoption of Generative AI, and increasingly stringent regulatory requirements. While existing initiatives address governance, compliance, and security in isolation, they provide limited guidance on how sovereignty can be operationalized at the architectural level.
In this paper, we argue that sovereignty must be treated as a first-class architectural property rather than a purely regulatory objective. We introduce a Sovereign Reference Architecture that integrates self-sovereign identity, blockchain-based trust and auditability, sovereign data governance, and Generative AI deployed under explicit architectural control. The architecture explicitly captures the dual role of Generative AI as both a source of governance risk and an enabler of compliance, accountability, and continuous assurance when properly constrained.
By framing sovereignty as an architectural quality attribute, our work bridges regulatory intent and concrete system design, offering a coherent foundation for building auditable, evolvable, and jurisdiction-aware AI-enabled systems. The proposed reference architecture provides a principled starting point for future research and practice at the intersection of software architecture, Generative AI, and digital sovereignty.
\end{abstract}


\maketitle

\section{Introduction}

Digital sovereignty has become a concrete and increasingly relevant challenge for software-intensive systems worldwide~\cite{tantoush2025balancing}, driven by concentrated cloud dependencies, rapid adoption of Generative AI, and stricter regulatory and assurance requirements.

Sovereignty, in the digital context, refers to the ability of a state, organization, or community to retain effective control, autonomy, and independent decision-making over its digital infrastructures, data, and technological processes. 
This control implies that systems, including the data they process, store, and exchange, can be designed, operated, evolved, and governed according to locally defined legal, ethical, operational, and societal principles, without undue reliance on external authorities, opaque mechanisms, or critical foreign dependencies~\cite{esposito_large_2025}. Digital sovereignty therefore goes beyond ownership or data location; it concerns who ultimately has the power to decide, enforce, and adapt how technology behaves over time~\cite{singi2020data}.

These sovereignty concerns manifest across different geopolitical and organizational contexts, including governments, critical infrastructures, and large enterprises seeking greater autonomy over their digital assets. The European Union represents one of the most structured and explicit responses to this challenge. In particular, the EU Cloud Sovereignty Framework~\cite{EUCloudSovereigntyFramework} articulates sovereignty as a multi-dimensional concept spanning strategic, legal, data, operational, supply-chain, technological, security, compliance, and sustainability dimensions. This framework makes explicit that sovereignty cannot be reduced to narrow requirements such as data localization or regulatory compliance alone (e.g., GDPR~\cite{GDPR}). Instead, it requires that digital systems operate under a clearly defined jurisdiction, avoid critical external dependencies, and provide transparent and auditable guarantees that support long-term autonomy, resilience, and accountability.
Prior software engineering research has shown that blockchain-based architectures can effectively support transparency, traceability, and governance in complex, multi-stakeholder systems, providing a technical foundation for autonomy and verifiable control \cite{marchesi_industrial_architecture_2022}.

In this position paper, we argue that addressing digital sovereignty requires a shift from ad hoc technical or regulatory measures toward a principled architectural response. Specifically, we contend that any actor aiming for meaningful digital sovereignty requires a \textbf{Sovereign Reference Architecture} (SRA) that treats sovereignty as a first-class architectural objective. Such an objective cannot be achieved through incremental refinements of existing cloud-centric practices alone. Instead, it must emerge from the deliberate integration of three \textbf{key architectural components}:

\begin{enumerate}
    \item \textbf{Self-Sovereign Identity (SSI)}, enabling decentralized and user-controlled digital identities;
    \item \textbf{Blockchain-based trust and auditability}, providing tamper-resistant provenance, accountability, and lifecycle verification across software and AI systems;
    \item \textbf{Sovereign AI infrastructures and models}, trained, deployed, and executed within jurisdiction-controlled cloud and edge environments.
\end{enumerate}


Our \textbf{goal} is to outline a coherent architectural perspective that brings these elements together and to highlight how the software architecture community can guide future research and practice toward the systematic realization of sovereign, trustworthy, and AI-enabled systems. The European context serves as a concrete instantiation of these ideas, but the proposed architectural principles are intended to be broadly applicable.

As a position paper, this work intentionally adopts a high-level architectural perspective, aiming to surface sovereignty-relevant concerns, constraints, and trade-offs rather than to provide a detailed component-level specification.

\section{Digital Sovereignty as an Architectural Principle}
While sovereignty is often framed as a policy or regulatory concern, it also constitutes a fundamental architectural property that affects system design decisions, quality attributes, and long-term maintainability.

From an SE perspective, sovereignty captures a combination of controllability, autonomy, and verifiability across the software stack.
In this paper, the term sovereign does not denote a new class of technologies, but rather the imposition of explicit architectural constraints on otherwise conventional components.
A component is considered sovereign insofar as its ownership, deployment, evolution, and verification remain under the control of the system stakeholders and within a clearly defined governance and jurisdictional boundary.

At the legal and governance layer, sovereignty means ensuring that system behavior, data flows, and operational processes remain within a predictable, enforceable governance boundary. For architects, this translates into constraints on where components may execute, how data is stored and transferred, and which external dependencies may be incorporated into the system.

At the operational and technological layer, sovereignty concerns the capability of system owners to maintain, modify, and evolve the system independently. Architectural decisions involving proprietary APIs, opaque cloud services, AI black boxes, or non-auditable components directly affect sovereignty by reducing observability, portability, or substitutability. The more a system relies on external, uncontrollable components, the lower its sovereignty.

At the supply-chain layer, sovereignty requires transparency and traceability across the software supply chain. As modern systems increasingly depend on third-party libraries, ML models, container images, CI/CD pipelines, and hardware accelerators, the architecture must support provenance tracking, integrity verification, and component validation. These capabilities are essential to prevent hidden dependencies or vulnerabilities from undermining system autonomy.

Empirical studies on blockchain-based systems show that insufficient provenance and lifecycle transparency directly contribute to governance and security risks, reinforcing the architectural relevance of tamper-resistant trust mechanisms \cite{marchesi_security_checklists_2025}.

At the data and AI layer, sovereignty involves full control over data lifecycle and AI lifecycle processes. This includes auditable and reproducible pipelines for data collection, preprocessing, model training, evaluation, deployment, and monitoring. Architecturally, this demands mechanisms that ensure traceability, explainability, and the ability to migrate or reimplement AI components without systemic lock-in.

Similar limitations have been observed in smart contract–based systems, where lack of architectural auditability has been shown to correlate with persistent security and governance vulnerabilities \cite{salzano_bridging_gap_2026}.

Viewing sovereignty as an architectural property compels software architects to rethink foundational aspects of system design, including trust boundaries, dependency management, deployment strategies, and runtime governance. 
Traditional cloud-centric architectures are insufficient: sovereignty must be embedded through verifiable constraints and design choices that preserve control, accountability, and evolvability.

\begin{figure}
    \centering
    \includegraphics[width=0.9\linewidth]{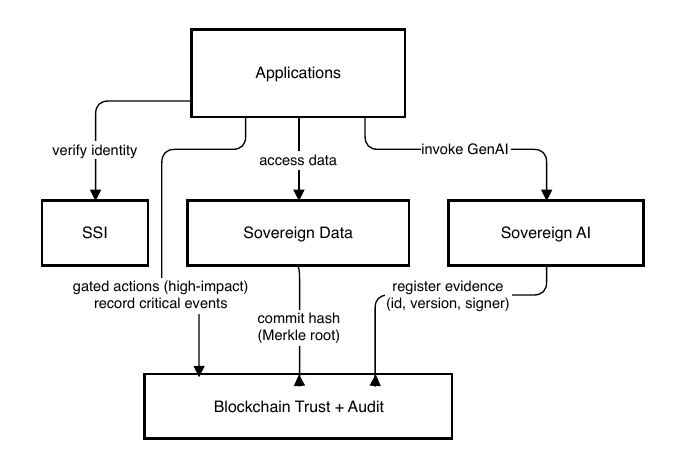}
    \caption{Simplified Overview of the Proposed Reference Architecture}
    \label{fig:sra}
\end{figure}

Blockchain acts as a cross-cutting trust substrate that anchors governance-relevant events across SSI, data, AI, and application layers.

\begin{table*}[t]
\caption{Layer contracts of the Sovereign Reference Architecture (SRA).}
\label{tab:sra-layer-contracts}
\centering
\small
\setlength{\tabcolsep}{3.2pt}
\renewcommand{\arraystretch}{1.4}
\begin{tabular}{p{2.5cm} p{2.75cm} p{5.8cm} p{5.6cm}}
\hline
\textbf{Layer} & \textbf{Purpose} & \textbf{Constraints} & \textbf{Mechanisms} \\
\hline

SSI &
Identity, authZ &
No foreign IdP lock-in; verifiable access; revocation &
DIDs; VCs; wallets; verifiers; revocation registry \\

Blockchain trust/audit &
Evidence, provenance &
Non-repudiation; tamper resistance; supply-chain traceability &
Hash/Merkle anchoring; signed attestations; audit events; provenance records \\

Sovereign data &
Govern + process data &
Residency; access control; retention/deletion; lineage &
Policy-as-code; local key mgmt; encryption; lineage capture; redaction gates \\

Sovereign AI &
Govern GenAI lifecycle &
Approved models only; reproducible eval; auditable use; leakage control &
Model registry; eval/promotion gates; prompt/tool evidence; sovereign telemetry \\

Applications &
Business services &
No policy bypass; controlled egress; sovereign observability; replaceability &
Policy enforcement point; API gateway; service mesh/mTLS; allowlists; audit hooks \\

\hline
\end{tabular}
\end{table*}

\section{A Sovereign Reference Architecture}
The proposed Sovereign Reference Architecture (SRA) is built on the principle that sovereignty emerges from architectural constraints that govern how components are created, deployed, operated, and evolved over time. Its layered design reflects the distinct requirements of identity governance, trust and provenance, data control, and AI lifecycle integrity.
Importantly, the SRA does not assume sovereignty to emerge from novel technological components, but from architectural constraints that govern how otherwise conventional components are integrated, controlled, audited, and evolved.

\subsection{Overview}
At its foundation, the SRA relies on sovereign cloud and edge infrastructures: computational and storage environments operated within the jurisdiction of interest and free from critical external dependencies that could compromise autonomy. Building on this foundation, the architecture integrates the following layers, each addressing a distinct sovereignty-critical concern:

\begin{itemize}
\item a \emph{Self-Sovereign Identity (SSI)} providing cryptographic, decentralized, and verifiable identities for individuals, organizations, devices, and AI agents, enabling identity assurance without reliance on centralized or foreign-controlled authorities;
\item a \emph{blockchain-based trust and audit} anchoring governance-relevant events and provenance across identity, data, AI, and supply-chain layers;

\item a \emph{sovereign data layer}, ensuring that data storage, processing, retention, and deletion remain under the governance, control, and jurisdiction intended by system stakeholders;
\item a \emph{sovereign AI layer}, enabling training, fine-tuning, evaluation, deployment, and inference of LLMs and multi-agent systems within controlled environments, and ensuring observability, reproducibility, and auditability of AI behaviour;
\item an \emph{application layer}, separating domain functionality from deployment, orchestration, and runtime governance concerns, enabling sovereignty constraints to be applied differently at the service composition and operational level.
\end{itemize}

While these concerns are presented as distinct architectural layers, blockchain-based trust mechanisms play a cross-cutting role within the SRA. They act as an architectural backbone that enforces sovereignty across identity, data, AI, and supply-chain layers, thereby transforming sovereignty from a declarative requirement into an enforceable and verifiable system property. This architectural role builds on prior software engineering frameworks for blockchain-based systems, where distributed ledgers were shown to act as foundational trust layers across heterogeneous components and organizational boundaries \cite{marchesi_ABCDE_2020}.

This generic reference architecture can be directly mapped onto the European context by instantiating its sovereignty constraints using the principles and assurance levels defined in the EU Cloud Sovereignty Framework, which specify how each architectural layer must operate under exclusive EU jurisdiction and governance.

Figure~\ref{fig:sra} illustrates the layered structure and sovereignty-relevant interactions of the SRA, whereas Table~\ref{tab:sra-layer-contracts} provides a consolidated view of its layers, associated purposes, architectural constraints, and enabling mechanisms.

\subsection{The role of blockchain}
Blockchain (or, more broadly, DLT) plays a key role in enabling sovereignty as a software quality attribute. Whereas traditional systems depend on centralized logs and trust anchors, sovereign systems require tamper resistance, transparency, decentralized verification, and durable audit trails.
Blockchain-based solutions have been successfully applied in regulated domains to guarantee end-to-end traceability, provenance, and auditability, demonstrating their suitability as trust substrates in sovereignty-critical architectures \cite{marchesi_agri_food_access_2022}.

From a software engineering perspective, blockchain contributes to sovereignty by:
\begin{itemize}
    \item \textbf{Reinforcing trust boundaries}, offering non-repudiable records of system actions, supply-chain artefacts, and AI lifecycle events;
    \item \textbf{Supporting provenance tracking}, enabling architects to verify component origins, model versions, training data dependencies, and artefact integrity;
    \item \textbf{Enforcing accountability}, ensuring that autonomous components, such as LLM-driven agents, leave verifiable traces of their decisions and interactions;
    \item \textbf{Strengthening identity infrastructures}, enabling decentralized identifiers, revocation registries, and cross-domain credential verification.
\end{itemize}

These capabilities make blockchain a foundational enabler for autonomy, auditability, and lifecycle assurance in complex software systems. The same capabilities support EU sovereignty objectives such as SOV-5 (Supply Chain) and SOV-7 (Security \& Compliance), and align with initiatives like EBSI~\cite{ebsi} and eIDAS2~\cite{eidas2}.

Without a distributed, tamper-resistant trust layer, key sovereignty properties such as accountability, provenance, and lifecycle assurance cannot be reliably enforced at system scale. In this sense, blockchain is not an optional infrastructure component, but a foundational architectural mechanism for operationalising digital sovereignty.

The SRA does not assume that all artefacts or events are recorded on-chain; blockchain anchors governance-relevant provenance. Cost and scalability concerns can be managed via permissioned and selective anchoring strategies.

\subsection{The Role of Generative AI}

Generative AI (GenAI), encompassing Large Language Models (LLMs) and LLM-driven agentic systems, is rapidly becoming a structural element of contemporary software architectures~\cite{esposito_genaimlr_2025}. Beyond isolated automation tasks, GAI increasingly participates in decision-making, coordination, and governance-related activities, thereby influencing architectural control, accountability, and sovereignty properties~\cite{esposito_2024_esem,esposito_beyond_2024}.

Within a sovereignty-oriented reference architecture, Generative AI must be treated as a first-class architectural concern. On the one hand, GAI introduces new sources of opacity, dependency, and governance risk; on the other hand, empirical evidence suggests that, when appropriately constrained, it can actively support governance, compliance, and assurance processes. This dual role positions Generative AI as both a problem to be managed and a mechanism through which sovereignty can be operationalized~\cite{esposito_beyond_2024}.

Within the SRA, GenAI-related risks are addressed at the architectural level by constraining model locality, lifecycle traceability, observability, and governance boundaries, rather than by relying on model-internal mechanisms alone.

\subsubsection{GenAI Risks and Architectural Mitigations}
GenAI introduces sovereignty-relevant risks, which the SRA addresses through architectural constraints and governance mechanisms:
\begin{itemize}
    \item \textbf{Opacity and accountability}: LLM-based components behave as black boxes, complicating traceability and responsibility attribution; the SRA mitigates this through observability requirements and governance-relevant logging~\cite{esposito_beyond_2024}.
    \item \textbf{External dependencies and data protection}: reliance on non-sovereign infrastructures and proprietary providers can reduce autonomy and increase leakage risks; the SRA constrains locality and lifecycle governance of data and models~\cite{esposito_large_2025}.
    \item \textbf{Operational governance}: as GenAI supports decision-making, continuous compliance becomes critical; within controlled environments, GenAI can assist risk analysis and documentation and enable continuous assurance via monitoring and anomaly detection, while preserving human-in-the-loop oversight~\cite{esposito_large_2025,esposito_beyond_2024}.
\end{itemize}

\subsubsection{Architectural Trade-offs and Implications}
In practice, sovereignty is frequently traded off against cost, scalability, and user convenience, rather than pursued as an absolute objective.
The integration of Generative AI within a sovereign reference architecture entails:
\begin{itemize}
    \item \textbf{Control vs. capability}: restricting GenAI to sovereign infrastructures may limit access to proprietary models but increases transparency and auditability.
    \item \textbf{Automation vs. accountability}: agentic systems can automate governance tasks, but require safeguards for traceability, reproducibility, and responsibility attribution.

\end{itemize}

To address these trade-offs, the SRA constrains where and how GenAI is trained and executed, enforcing locality, lifecycle traceability, and behavioural observability.

These considerations align with obligations defined in the EU AI Act for high-risk systems and with SOV-3 (Data and AI Sovereignty).

\section{Challenges and Future Directions}

Achieving sovereignty as a system-level architectural property raises several challenges. Architecturally sovereign systems must address hardware dependencies (e.g., accelerators and secure enclaves), cloud portability, verifiable software supply chains, and auditable AI behaviour. Moreover, architectural support for sovereign identity management, verifiable trust, and runtime compliance monitoring remains an open challenge, especially in large-scale, distributed, and AI-intensive systems.

From a research perspective, these challenges highlight the need for architecture-level patterns and decision models that explicitly support sovereign microservices and AI-native systems. This includes methods to audit LLM and multi-agent behaviours, ensure provenance and lifecycle traceability, and integrate mechanisms such as quantum-safe trust infrastructures.

These issues mirror ongoing discussions in European initiatives such as Gaia-X, NIS2, the AI Act, and emerging cybersecurity certification schemes. While frameworks such as Gaia-X emphasize governance and data sovereignty~\cite{gaiax}, NIS2 and the AI Act focus on organizational accountability and risk management~\cite{nis2,aiact}. In parallel, EUCS targets assurance levels and security requirements~\cite{eucs}, while international standards target interoperability and portability~\cite{iso19941}. As a result, software architects face heterogeneous requirements that remain only loosely connected to architectural decision-making.

Empirical evidence suggests that continuous assurance is essential as systems become increasingly autonomous and distributed~\cite{lunesu_agile_risk_2021}. In this context, the proposed SRA acts as a connective layer between regulatory intent and architectural realization, translating sovereignty, security, and accountability requirements into architectural constraints and design decisions, while supporting interoperability across federated cloud and data spaces~\cite{idsa}.

\section{Conclusions}
This paper positions sovereignty as a core architectural concern for modern software-intensive systems. The proposed SRA integrates SSI, blockchain-based trust mechanisms, sovereign data governance, and sovereignty-preserving AI into a unified architectural structure. Rather than presenting sovereignty as a universally optimal solution, the SRA provides an architectural lens for making sovereignty-related trade-offs explicit, auditable, and governable. The SRA aligns naturally with the EU Cloud Sovereignty Framework, providing a conceptual and architectural foundation for addressing sovereignty objectives (SOV-1 to SOV-8) and achieving higher SEAL assurance levels.
Finally, our results show that blockchain does not replace traditional audits, but serves as an architectural trust substrate when sovereignty must span autonomous actors and jurisdictions where centralized audit assumptions break down.

\bibliographystyle{ACM-Reference-Format} 
\bibliography{main}
\end{document}